\documentstyle[prd,eqsecnum,preprint,tighten,aps]{revtex}
\begin{document}
\draft
\title{Casimir energy of a non-uniform string}
\author{L.~Hadasz\thanks{Electronic address: hadasz@thrisc.if.uj.edu.pl}}
\address{Jagiellonian University, M. Smoluchowski Institute of Physics,
Reymonta 4, 30--069 \\ Cracow, Poland}
\author{G.~Lambiase\thanks{Electronic address: lambiase@physics.unisa.it}}
\address{Dipartimento di Scienze Fisiche ``E.R. Caianiello'',
 Universit\'a di Salerno, 84081 \\ Baronissi (SA), Italy}
\author{V.\ V.~Nesterenko\thanks{Electronic address: nestr@thsun1.jinr.ru}}
\address{Bogoliubov Laboratory of Theoretical Physics,
Joint Institute for Nuclear Research, 141980 Dubna,  Russia}
\date{\today}
\maketitle
\begin{abstract}
The Casimir energy of a non-uniform string built up from two
pieces with different speed of sound is calculated. A standard
procedure of subtracting the energy of an infinite uniform string
is applied, the subtraction being interpreted as the
renormalization of the string tension. It is shown that in the
case of  a homogeneous string this method is completely equivalent
to the zeta renormalization.
\end{abstract}

\section{Introduction}
In the standard setting of the Casimir effect problem, a bounded
 configuration space of
the quantum field system is considered~\cite{PMG,MTr}. However
unbounded configuration space can be nonhomogeneous, i.e., it can
consist from separate regions endowed with different physical
characteristics. In the case of electromagnetic field such
caracteristics are permittivity $\varepsilon $ and permiability
$\mu $ of the media. It is obvious that the vacuum energy  should
depend on the configuration of these nonhomogenities. Hence the
problem on finding out this dependence, i.e.\ on calculation of
the Casimir energy, can be set.

When calculating the vacuum energy it proves to be essential
whether the velocity of light (in the general case the velocity of
relevant quanta) is continuous when crossing the interface between
regions with different characteristics. From a mathematical stand
point the constant velocity of light implies that the coefficients
of the higher (usually the second) derivatives in the
corresponding dynamical equation are continuous functions in
space. If it is the case then the calculation of the Casimir
energy (or the Casimir forces) turns out to be practically  the
same as for an empty space with a perfectly conducting shell
having the shape of the interface between
media~\cite{BK,BN,MNN,BNP,NP1,NP2}.

The discontinuity of the speed of light on the boundaries between different
regions results in considerable mathematical difficulties. Certain results have
been obtained here only recently. The Casimir energy was calculated
 for a compact dilute dielectric ball proceeding from the Green's function
formalism and employing the ``naive'' zeta function technique for
removal of the divergences without detailed procedure of analytic
continuation~\cite{BMM,BM,Barton,Mar}. It was assumed  that the
difference of the light velocities inside ($c_1$) and outside
($c_2$) is small and only the term proportional to $(c_1 - c_2)^2$
in the Casimir energy has been kept. In Ref.~\cite{Bordag} it was
shown that the heat kernel coefficient $a_2$,  responsible for the
pole contribution to the Casimir energy, vanishes in this
approximation (it proves to be proportional to $(c_1-c_2)^3$).
This implies that the complete zeta regularization should provide
a finite answer in this problem.

In calculations accomplished in Refs.~\cite{BMM,BM,Mar} a key role
was played by
 the so
called contact terms which cannot be incorporated in the standard
zeta function technique at least in a straightforward way. Their
physical origin is still unclear.

Thus it is worth investigating  these problems in the framework of
a more simple model. Such a model, preserving all essential
features  of  field theory, is a string built up from two pieces
which have  different velocities of sound (the velocities of
vibration propagation). In papers by I.\ 
Brevik and co-authorths\ 
\cite{BNil,BBO,BESA} a piecewise
uniform string, with the same velocity of
sound at each of its constituents,
 has been studied. 
 Such string model  is a one-dimensional
analog of the electromagnetic field in media obeying the condition
$\varepsilon \mu =c^{-2}$, where $c$ is the constant velocity of
light. In this model the Casimir energy has been investigated in
detail both at zero and finite temperature.

The present paper seeks to calculate the Casimir energy of a
nonuniform open string of total length $R$, which is built up of
the pieces of length $r$ and $R-r$ having the sound velocities
$v_1$ and $v_2$ respectively.
 At the joint point ($x=r$) the continuity of the string shape
and its tangents is imposed. The string ends $x=0$ and $x=R$ are
subjected to the Dirichlet or Neumann boundary conditions. The
equations defining the eigenfrequencies of the string are derived.
This material is presented in Sect.\ II.  Proceeding from this in
Sect.\ III the Casimir energy of the string is calculated by
making use of the mode-by-mode summation method~\cite{NP1,MNN}, which relies
 on the
contour integration. The renormalization of the string tension by
subtracting the energy of an infinite uniform string enables one
to remove
 the divergences in a unique way. It is shown that in the  limit $R \to
 \infty $ the Casimir energy of the string does not depend on the boundary
conditions imposed at $x=R$ and is expressed in terms of the
polylogarithm function.  In the Conclusion (Sect.\ IV) the results
obtained are discussed briefly. In the Appendix  it is shown that the
subtraction procedure employed in the paper is equivalent to the
zeta renormalization  when considering a  usual uniform string.
\section{Classical dynamics of a non-uniform string}
The model we are considering
 is a string built up from two pieces of length $r$ and $R-r$ and endowed
with the sound velocities $v_1$ and $v_2$, respectively. The string is
stretched along the $X^1$ axis (which we shall denote by $x$ in what follows)
and can vibrate in $D-2$ transverse dimensions $X^2\ldots X^D$.
The  both pieces of the string satisfy the wave equation
\begin{equation}
\label{waveeq}
\frac{1}{v_j^2}\frac{\partial^2X^i}{\partial t^2} -
\frac{\partial^2X^i}{\partial x^2} = 0,
\quad j= 1,2, \quad i=2,3,\dots ,D\,{.}
\end{equation}
At the ends the string is subjected to the boundary conditions, which
we shall take to be of either Dirichlet or Neumann type. Consequently,
we shall consider four cases:
\begin{eqnarray}
\label{bcond}
  X^i(t,0) &=& 0, \quad  X^i(t,R) = 0, \quad \text{(DD)}, \nonumber \\
  X^i(t,0)& = & 0,  \quad
\left.\frac{\partial X^i}{\partial x}\right|_{x=R} =0, \quad
\text{(DN)}, \nonumber \\
  \left.\frac{\partial X^i}{\partial x}\right|_{x=0}& =&0,
\quad  X^i(t,R) = 0, \quad \text{(ND)},         \nonumber  \\
 \left.\frac{\partial X^i}{\partial x}\right|_{x=0} &=&0, \quad
\left.\frac{\partial X^i}{\partial x}\right|_{x=R} = 0 \quad
\text{(NN)},        \nonumber \\
&& i=2,\ldots , D\,{.}
\end{eqnarray}
At the joining point $x=r$ we impose the continuity conditions
\begin{eqnarray}
\label{joining}
\lim_{x\to r_{-}}X^i(t,x) & = & \lim_{x\to r_{+}}X^i(t,x), \nonumber \\
\lim_{x\to r_{-}}\left(\frac{\partial X^i}{\partial x}\right) & = &
\lim_{x\to r_{+}}\left(\frac{\partial X^i}{\partial x}\right),
\quad  i = 2,\ldots, D.
\end{eqnarray}

The wave equations (\ref{waveeq}) together with the boundary
conditions (\ref{bcond}) and the continuity  conditions
(\ref{joining}) lead to the equations which determine the
eigenfrequencies of the string~$\omega $
\begin{eqnarray}
\label{fs}
f_1(\omega ) & =& \sin(\alpha\omega) + \delta v \, \sin(\beta\omega) = 0,
\quad \mbox{(DD)},\nonumber \\
f_2(\omega) & =&
\cos(\alpha\omega) - \delta v \, \cos(\beta\omega) = 0, \quad \mbox{(DN)},
\nonumber  \\
f_3(\omega) &  =&
\cos(\alpha\omega) + \delta v \, \cos(\beta\omega) = 0, \quad
\mbox{(ND)}, \nonumber \\
f_4(\omega ) &=&
\sin(\alpha\omega) - \delta v \, \sin(\beta\omega) = 0,\quad \mbox{(NN)},
\end{eqnarray}
where
$$
\delta v = \frac{v_1-v_2}{v_1+v_2}, \quad
\alpha = \frac{r}{v_1} + \frac{R-r}{v_2},
\quad
\mbox{and}
\quad
\beta = \frac{r}{v_1} - \frac{R-r}{v_2}{.}
$$
Let us note, that $\alpha > 0$ and $\alpha > \beta$.
\section{Ground state energy of the string
and its renormalization} When calculating the Casimir energy, the
main problem encountered here is the removal of the divergences,
which arise  inevitably in such calculations.
 In quantum field
theory~\cite{BogSh} this procedure is accomplished in the course of the
renormalization of the physical parameters, which specify the
field theory model under investigation (the masses of the quanta,
coupling constants an so on). In order for this procedure to be
correct from mathematical stand point, divergent expressions
should be regularized, the regularization being removed after
renormalization.

         We define the renormalized Casimir energy in a standard way~\cite{PMG}
\begin{equation}
\label{defE}
E_{\text{C}}=\frac{1}{2}\sum_n\left (
\omega _n - \overline{\omega}_n
\right ),
\end{equation}
where the $\omega $'s are the eigenfrequencies of the string,
i.e., the roots of the frequency equations~(\ref{bcond}), and the
$\overline{\omega } $'s are the roots of the same equations but in
the limit when the total string length $R$ and $r$ tend to
infinity, and the difference of the velocities $\delta v$ tends to
zero. Hence in this limit the initial string becomes a uniform
infinite string. In quantum field theory this implies the removal
of the Minkowski space contribution~\cite{PMG}. In the Appendix
it is shown that the method of calculation of the vacuum energy,
outlined above, is completely equivalent to the zeta
regularization in the case of usual uniform string.

It is convenient to represent the sum (\ref{defE}) in terms of the
contour integral~\cite{WW}
\begin{equation}
\label{oint}
E_{\text{C}}=\frac{D-2}{4 \pi i} \oint_C dz\,z\,\frac{d}{dz}\ln\frac{f(z)}
{\overline{f}(z)}\,{,}
\end{equation}
where the function $f(z)$ is any function $f_l(z), \quad l=1, \dots , 4$
 defining the frequency equations in~(\ref{fs}), and $\overline{f} $ is obtained from
$f$ when passing to the limit described above. The contour $C$
encloses all the positive roots of the equation $f(\omega)=0$.
This contour can be deformed into the semicircle $C_\Lambda $  of
radius $\Lambda $ in the right half-plane and the
interval of the imaginary axes $(-i\Lambda, i\Lambda )$. When the
radius $\Lambda $ is fixed, the contour $C$ encloses a finite
number of the roots of the equation $f(z)=0$. The sum of these
roots is obviously finite. Hence the radius $\Lambda $ is a
regularization parameter, and taking the limit $\Lambda \to \infty
$ means the removal of the regularization.

The subtraction  in Eq.\ (\ref{oint}) accomplished with the
function $\overline{f}(z)$ can be interpreted as the
renormalization of the parameters which specify the classical
energy of the string. From the general consideration one can
assume that this energy is proportional to the total length of the
string $R$, i.e., $E^{\text{cl}}=T\cdot R$, where $T$ is a
dimensional parameter $[T]=L^{-2}$. It is determined by a concrete
string model that should certainly involve nonlinearities. For
example, in the relativistic string model~\cite{BarN} this
parameter is the string tension. Due to the quantum corrections
the string tension should be renormalized in the following way
\begin{equation}
\label{tension}
T^{\text{ren}}=T+\frac{D-2}{R}\frac{1}{4\pi i} \oint _C dz\,z\,\frac{d}{dz}\ln
\overline{f}(z).
\end{equation}
After removing the regularization $(\Lambda \to \infty)$ the integral in
(\ref{tension})        obviously  diverges. However it is assumed that this
divergence is canceled by the respective infinity of the ``bare'' string
tension $T$. As a result the renormalized (physical) string tension
 $T^{\text{ren}}$
proves to be finite. Here we follow the standard procedure of
removing the divergences in quantum field theory~\cite{BogSh}. In
view of the oscillating behavior of the  functions $f_l(z), \quad
l=1, \dots, 4$ on the semicircle $C_\Lambda $ their asymptotics
for $R\to \infty $ are simply these functions themselves, i.e.,
$\overline{f}_l(z)=f_l(z),\quad z\in C_\Lambda $. Therefore the
integration along this part of the contour $C$ does not give
contribution into the Casimir energy (\ref{oint}). On the
imaginary axes the asymptotics of the functions $f_l(z)$ when
$R,\;r \to\infty $ can be find easy
\begin{equation}
f_l(z)\to \frac{e^{\alpha y}}{2},\quad \alpha =\frac{r}{v_1}+
\frac{R-r}{v_2}\,{.}
\end{equation}
As a result we obtain the final expression for the Casimir energy
\begin{equation}
\label{Efinal}
E_{\text C}=\frac{D-2}{2 \pi}\int_0^\infty dy\ln{[h_l(y)]}  ,
\quad l=1,\dots , 4
\end{equation}
with the functions $h_l(y)$ given by
\begin{eqnarray}
h_1(y) & = & 1 - {\rm e}^{-2\alpha y} +
    2\delta v\ {\rm e}^{-\alpha y}\sinh(\beta y), \quad \mbox{(DD)},
\nonumber\\
h_2(y) & = & 1 + {\rm e}^{-2\alpha y} -
    2\delta v\ {\rm e}^{-\alpha y}\cosh(\beta y), \quad \mbox{(DN)},
\nonumber\\
h_3(y) & = & 1 + {\rm e}^{-2\alpha y} +
    2\delta v\ {\rm e}^{-\alpha y}\cosh(\beta y), \quad \mbox{(ND)},
\nonumber\\
h_4(y) & = & 1 - {\rm e}^{-2\alpha y} -
    2\delta v\ {\rm e}^{-\alpha y}\sinh(\beta y), \quad \mbox{(NN)}.
\label{hs}
\end{eqnarray}

As a simple consistency check, we  apply the formulas
(\ref{Efinal}) and (\ref{hs}) to the uniform string ($v_1=v_2
\equiv v$) of length $R$ with the expected
 result
\cite{lusher}
\[
E_{\text{C}} =\frac{D-2}{2\pi}\int_0^\infty dy \ln \left(
1- e^{-2\frac{R}{v}y}\right )
= -\frac{\pi(D-2)}{24R}v
\]
in the DD and NN cases and
\[
E_{\text{C}} =\frac{D-2}{2\pi}\int_0^\infty dy \ln \left(
1+ e^{-2\frac{R}{v}y}\right )
= \frac{\pi(D-2)}{48R}v
\]
in the DN and ND cases.

Let us consider the string configuration, when the second piece of the string
becomes infinitely long (i.e. $R$ tends
to infinity, while $r$ remains fixed). Then we have
\begin{eqnarray}
h_1^{\infty}(y) & =& h_2^{\infty}(y) =
1-\delta v\ {\rm e}^{-2\frac{r}{v_1}y} \equiv h_-(y),
 \quad \mbox{(DD)},\;\mbox{(DN)},   \nonumber  \\
h_3^{\infty}(y) & = &h_4^{\infty}(y) =
1+\delta v\ {\rm e}^{-2\frac{r}{v_1}y} \equiv h_+(y), \quad \mbox{(ND)},\;
\mbox{(NN)}.
\label{hs2}
\end{eqnarray}
These formulas imply that when $R\to \infty $
 the Casimir energy becomes independent of the type
of the boundary condition imposed at  $x=R$. Certainly, this is an
appealing property of the vacuum energy of the string under
consideration.
 For $R\to \infty$ it is also possible to obtain
the explicit form of this  energy
\begin{eqnarray}
E_\pm &=& \frac{D-2}{2\pi }\int_0^\infty \!\!\! dy
\ln \left [h_\pm(y) \right ] \nonumber
\\
      &=&
    \frac{D-2}{2\pi}
    \int_0^{\infty}\!\!\!dy
    \ln\left( 1\pm\delta v\ {\rm e}^{-\frac{2ry}{v_1}}\right ) \nonumber  \\
       &=&     -    \frac{D-2}{4\pi}
\,\frac{v_1}{r} \, \text{Li}_{2}(\pm \delta v)\,{,}
\label{poly}
\end{eqnarray}
where
\begin{eqnarray*}
\text{Li}_\nu(z)&= &  \sum_{k=1}^{\infty}\frac{x^k}{k^\nu }, \quad |z|<1,
\nonumber \\
&=&\frac{z}{\Gamma (\nu)}\int _0^\infty \frac{t^{\nu-1}dt}{e^t-z}, \quad \Re \,
\nu >0, \quad  |\text{arg} (1-z)|<\pi
\end{eqnarray*}
is the polylogarithm function \cite{erdelyi}.

Our consideration of non-uniform string can be generalized in a
straightforward way to the finite temperature. It can be done by
the substitution~\cite{NP3}
\begin{equation}
dy \to 2\pi \theta \mathop{{\sum}'}\limits_{n=0}^{\infty} \delta
(y-\Omega_n)\,dy\,{,}
\end{equation}
where $\theta $ is the temperature, $\Omega _n=2 \pi n \theta$ are the Matsubara
frequencies, and the prime means that the term $n=0$ is
taken with half weight and
\section{Conclusion}
The non-uniform string with finite $r$ and $R\to \infty$  is, in
some sense, an analogue of the radial part of the problem
concerning the calculation of the vacuum energy of a pure
dielectric compact ball placed in a homogeneous unbounded medium
or cavity in such a medium \cite{BNP,BMM}.
 The main lesson of our consideration
is the following. When defining the counter term in Eq.\
(\ref{defE}), in order to renormalize the Casimir energy, we take
not only the limit $R,\,r\to \infty $ but also put $\delta v \to
0$. It means that all the physical parameters, specifying the
problem under study, should be involved in determination of the counter
term.

It is worth noting  that we do not use here the contact terms in
 order to get a finite
result for the Casimir energy.
\acknowledgments

This work has been  accomplished during the visits of V.V.N. to
the Jagellonian University (Cracow) and to the Salerno University.
It is a pleasure for him to thank Professors H.\ Arodz, G.\
Scarpetta, Drs. L.\ Hadasz, and G.\ Lambiase for warm hospitality.
The financial support of the Bogoliubov--Infeld program, IIASS
and INFN
is acknowledged.
\appendix
\section*{Riemann zeta function in the uniform string model}
The representation of the spectral zeta functions in terms of
contour integrals, with the integrands being the relevant frequency
equations,
 is of wide use~\cite{Bordag,LNB}  This representation
 is a direct application of the principle
of argument theorem from the complex analysis~\cite{WW,T}. However
 this theorem alone does not afford the required analytical
continuation of the zeta function. The details of this technique
can be clearly demonstrated by considering the zeta function for
the usual uniform string. In this case we are dealing with the
Riemann zeta function the analytic continuation of which is well
known~\cite{WW,T}.

The eigenfrequencies of the uniform string of the length $R$ with
the Dirichlet boundary conditions
\begin{equation}
\label{roots}
\omega_n=\frac{n\pi }{R}, \quad n=1,2,\dots
\end{equation}
are the positive roots of the equation
\begin{equation}
\label{sin}
\sin (\omega R)=0.
\end{equation}
For simplicity the velocity $v$ is taken to be~1. Let us define the zeta
function in the problem at hand by  the standard formula
\begin{equation}
\label{defz}
\zeta (s)=\sum _{n=1}^\infty \omega _n^{-s}, \quad  \Re\, s> 1.
\end{equation}
When $R=\pi$ we obviously have  the Riemann zeta function.

Now we represent the sum (\ref{defz}) in terms of the contour integral
\begin{equation}
\label{zcontour}
\zeta (s)=\frac{1}{2\pi i}\oint_Cz^{-s}\frac{d}{dz}\ln [\sin (zR)] dz\,{,}
\end{equation}
where the contour $C$ encloses  all the positive roots  of
Eq.~(\ref{sin}). The point $z=0$ is the branching point of the
function $z^{-s}= \exp(-s\ln z)$ in Eq.~(\ref{zcontour}).
Obviously this point should be left outside the closed contour $C$
in order to integrate one valued function. For the logarithm we
take as usual the branch acquiring real values  on the positive
half-axes $0<\Re\, z < \infty $. Let us deform (still formally)
the contour $C$ in a semicircle of infinitely large radius laying
in the right half-plane of the complex variable $z$ and close it
by imaginary axes $-\infty < \Im \, z < \infty $. For $ \Re \, s >
1 $ the integration along the semicircle can be dropt. As a result
one gets
\begin{eqnarray}
\zeta (s) &=&  \frac{1}{\pi} \sin\left ( \frac {\pi s}{2}\right )
\int_0^\infty dy\, y^{-s}\frac{d}{dy}\ln [\sinh (yR)]           \label{zeta1}
\\
&=& \frac{R}{\pi }  \sin\left ( \frac {\pi s}{2} \right )\int _0^\infty dy\,
y^{-s}\coth(yR)\,{.}\label{zeta2}
\end{eqnarray}
The integral in Eq.~(\ref{zeta2}) converges at the upper limit only for $\Re \,s
>1$ and at the lower limit for $\Re \;s<0$. The latter is due to the singular
behaviour   $\sim \varepsilon ^{-s}$ of the integral
 along the semicircle of infinitely small radius
$\varepsilon $ at the origin. This contribution
vanishes only for $\Re \,s <0$. Thus, in the general case the contour $C$ in
Eq.~(\ref{zcontour}) cannot pass through the point $z=0$.

However this contour can be retained if we take, instead of
Eq.~(\ref{sin}), a new frequency equation
\begin{equation}
\label{sin1}
\frac{\sin (\omega R)}{\omega R}=0\,{,}
\end{equation}
which has the same positive roots as Eq.~(\ref{sin}), but the point $z=0$ does
not satisfy it. As a result we obtain, instead of Eq.\  (\ref{zeta2}),
\begin{equation}
\label{zetacthmin}
\zeta (s) =\frac{R}{\pi }\sin \left (
\frac{\pi s}{2}
\right )       \int _0^\infty dy\, y^{-s}\left (
\coth (yR)-\frac{1}{yR}
\right )\,{.}
\end{equation}
This integral representation of the zeta function is defined for $1<\Re
s<2$ and gives in fact the same sum (\ref{defz}) but with explicit
substitution of the roots (\ref{roots}). Really, let us present the function
$\coth (y R)$ in Eq.~(\ref{zetacthmin}) as a series (see, for example,
Eq.~1.4.21.4 in~\cite{GR})
\begin{equation}
             \coth z=\frac{1}{z}+ 2z \sum_{n=1}^\infty \frac{1}{z^2+\pi
^2n^2}\,{.}
\end{equation}
After that integration in (\ref{zetacthmin}) can be done
\begin{eqnarray}
\label{zeta3}
             \zeta (s) &=& \frac{R}{\pi } \sin \left (\frac{\pi s}{2}
\right )
2R\sum_{n=1}^\infty \int _0^\infty \frac{y^{1-s}dy}{y^2R^2+\pi ^2 n^2}
\nonumber \\
&=&\frac{R}{\pi }\sin \left (\frac{2 \pi s}{2} \right )
\Gamma \left (\frac{s}{2} \right )
\Gamma \left (1 -\frac{s}{2} \right )\frac{1}{R} \sum_{n=1}^\infty
\left         (
\frac{n \pi}{R} \right )^{-s} \nonumber \\
&=&  \sum_{n=1}^\infty
\left         (
\frac{n \pi}{R} \right )^{-s}{.}
\end{eqnarray}
Thus  the contour integration of the frequency equation
(\ref{sin}) or (\ref{sin1}) gives us nothing new as compared with
the sum (\ref{defz}) with explicit frequencies~(\ref{roots}).

As known~\cite{WW,T}
analytic continuation of the series (\ref{defz}) with $R=\pi $
to the region $\Re s < 1$ is provided by the formula
\begin{equation}
\label{analytic}
\zeta_{\text{R}}(s)=\frac{i\Gamma (1-s)}{2  \pi} \int_\infty^{(0+)}\frac{(-z)^{s-1}}
{e^z-1}dz\,{.}
\end{equation}
Here $\zeta_{\text{R}}(z)$ is the Riemann zeta function,
 and the path of integration
encircles the real positive axes $0\le \Re\,z\le \infty$. When $\Re \,s>1$ the
integration contour around the origin can be drawn to the point, and the
remaining
integral gives
\begin{equation}
\zeta_{\text{R}}(s)=\frac{1}{\Gamma (z)}\int_0^\infty \frac{x^{s-1}}{e^x-1}dx
\,{.}
\end{equation}
Expanding the function $1/(e^x-1)$ in powers of $e^x$ and
 integrating  each term we
arrive at the sum
\begin{equation}
\label{zeta4}
\zeta _{\text{R}}=\sum_{n=1}^\infty \frac{1}{n^s}, \quad \Re \,s >1\,{.}
\end{equation}

If $\Re\, s< 1$, the integral around the origin in Eq.\ (\ref{analytic})
does not vanish, and it should be
taken into account when reducing the contour integral representation
(\ref{analytic}) to the ordinary  integral (see, for example, \cite{T}).

The Riemann  zeta function defined in the entire plane
 $s$, safe for the point $s=1$, by the contour integral (\ref{analytic})
obeys the Riemann reflection formula~\cite{WW,T}
\begin{equation}
\label{reflection}
\zeta_{\text{R}}(s) = \frac{(2\pi )^s}{\pi }\sin \left (
\frac{\pi s}{2}
\right )\Gamma (1-s) \zeta_{\text{R}}(1-s)\,{.}
\end{equation}
This formula relates the values of the zeta function left and
right from the line $\Re \,s =1$. At the point $s=1$ the function
$\zeta _{\text{R}}(s)$ has a simple pole
\begin{equation}
\left .\zeta_{\text{R}}(s)\right |_{s\to 1}\simeq \frac{1}{s-1}+\gamma\,{,}
\end{equation}
where $\gamma $ is the Euler constant. It is this pole that prevents
the use of the definition (\ref{zeta4}) for $\Re \,s <1$.

Thus we see that even for analytic continuation of the simple sum (\ref{zeta4})
a special choice of the path and the integrand is
needed.  Direct summation of the frequencies by contour integration does not
afford analytic continuation.

However if we define the zeta function for a uniform string by
subtracting the vacuum energy of an infinite string from
Eq.~(\ref{zeta4}),  as it is usually done when  calculating the
Casimir energy (see Eq.~(\ref{defE})), then the resulting zeta
function proves to be the analytic function for $\Re \, s <0$ and
exactly that function which follows from Eq.~(\ref{analytic}) or
from the reflection formula (\ref{reflection}), that is the same.

So let us define the zeta function in the following way
\begin{equation}
\label{zetanew}
\zeta(s)=\sum_n^\infty \left ( \omega _n^{-s}-\bar {\omega} _n^{-s}
\right )\,{,}
\end{equation}
where  the frequencies $\overline{\omega} $ are defined by Eq.\
(\ref{sin}) when $R\to \infty $. The corresponding limiting form
of this equation can be found unambiguously upon rotating the
integration path to the imaginary axes
\begin{equation}
\label{limit}
\lim_{R\to \infty} \sinh (y R) =\frac{e^{yR}}{2}\,{.}
\end{equation}
Now we apply  Eq.~(\ref{zeta1}) to the both terms in definition
(\ref{zetanew})
\begin{equation}
\label{zeta5}
  \zeta(s)=\frac{s}{\pi }\sin \left (
\frac{\pi s}{2}
\right )
\int_0^\infty y^{-s-1} \ln \left (
       1- e^{-2yR}
\right )dy \,{.}
\end{equation}
Here the integration by parts has been done, that is legitimated
for $\Re \,s <0 $. Expanding the logarithm in (\ref{zeta5}) in
power series and integrating each term we obtain for $\Re \, s<0 $
\begin{eqnarray}
\label{zetafinal}
\zeta (s) &=& \frac {(2\pi )^s}{\pi}\sin \left (
\frac{\pi s}{2}
\right )                 \Gamma (1-s)\sum_{n=1}^\infty \frac{1}{n^{1-s}}
\nonumber \\
&=& \frac {(2\pi )^s}{\pi}\sin \left (
\frac{\pi s}{2}
\right )                 \Gamma (1-s)\zeta(1-s)\,{.}
\end{eqnarray}
It is exactly the same result that follows from the Riemann reflection formula
(\ref{reflection}).

Thus the analytic continuation of the Riemann zeta function
(\ref{zeta4}) into the region $\Re \,s<1$, which is given
by Eq.\ (\ref{analytic}), implies in essence that for $\Re\, s <1$
one considers the initial series (\ref{zeta4}) with subtraction
(\ref{zetanew}). Certainly for $\Re \,s>1$
 the new definition
of the zeta function (\ref{zetanew}) cannot be reduced to the
initial  series (\ref{zeta4}), hence it cannot play the role of
Eq.\ (\ref{analytic}), i.e., it cannot afford, in a rigorous
mathematical sense,  analytic continuation
 of the series (\ref{zeta4}) to the region
$\Re \, s<1$.                  The new definition (\ref{zetanew}) simply
 ``guesses''  the result of analytic continuation of the Rieman zeta
function (\ref{zeta4}) to this region.

To our opinion, this consideration shows, in a  simple and clear
way, the relationship between the analytic continuation, providing
the mathematical basis of the zeta renormalization technique, and
the method  of subtraction or counter terms, widely used by
physicists.

Summarizing we arrive at the conclusion that the
 calculation of the Casimir energy
of a non-uniform string in  the  present paper is completely equivalent
to the zeta regularization in the case of a usual uniform string.


%
%

%
%

%

\end{document}